\pdfoutput=1
\documentclass[a4paper,12pt]{article}
\usepackage[english]{babel}

\usepackage{amsmath, amssymb, amsfonts, amsthm}
\usepackage{fancyvrb}
\usepackage{graphicx}
\usepackage{diagbox}
\usepackage{booktabs}

\usepackage{transparent}

\usepackage{bm}

\newcommand{\Ex}[1]{\mathbb{E}\!\left[{#1}\right]}

\usepackage{color}
\usepackage[dvipsnames]{xcolor}

\VerbatimFootnotes

\newcommand*{\tablefont}{\fontfamily{ppl}\selectfont}
\usepackage{dsfont}

\usepackage{a4wide}
\usepackage{textcomp}
\usepackage{setspace}
\usepackage[pagestyles]{titlesec}

\usepackage[hyperfootnotes=false]{hyperref}

\usepackage{color}

\VerbatimFootnotes

\usepackage{etoolbox}
\patchcmd{\equation}{\alph{section}}{.\arabic{equation}}{}{}

\usepackage{bbm}
\usepackage{caption}
\usepackage{multirow}
\usepackage{subcaption}
\usepackage{enumerate}
\usepackage{enumitem}
\usepackage{chngcntr}
\usepackage{mwe}
\usepackage{dsfont}
\usepackage{pdflscape}
\usepackage{lscape} 
\usepackage{epstopdf} 
\usepackage{verbatim}
\usepackage{tabu}
\usepackage{listings}
\usepackage{multirow}
\usepackage{xcolor}
\usepackage{bigints}
\usepackage[T1]{fontenc}
\usepackage{csquotes}

\makeatletter
\newcommand{\specialcell}[1]{\ifmeasuring@#1\else\omit$\displaystyle#1$\ignorespaces\fi}
\makeatother
\usepackage[margin=0.95in]{geometry}

\def\mydefgreek#1{\expandafter\def\csname b#1\endcsname{\text{\boldmath$\mathbf{\csname #1\endcsname}$}}}
\def\mydefallgreek#1{\ifx\mydefallgreek#1\else\mydefgreek{#1}%
	\lowercase{\mydefgreek{#1}}\expandafter\mydefallgreek\fi}
\mydefallgreek {beta}{Gamma}{Delta}{epsilon}{Omega}{etaex}{Theta}{Iota}{Lambda}{kappa}{mu}{nu}{Xi}{Pi}{rho}{Sigma}{tau}\mydefallgreek

\def\mydefb#1{\expandafter\def\csname b#1\endcsname{\mathbb{#1}}}
\def\mydefallb#1{\ifx#1\mydefallb\else\mydefb#1\expandafter\mydefallb\fi}
\mydefallb ABCDEFGHIJKLMNOPQRSTUVWXYZeucsxht\mydefallb

\def\mydefc#1{\expandafter\def\csname c#1\endcsname{\mathcal{#1}}}
\def\mydefallc#1{\ifx#1\mydefallc\else\mydefc#1\expandafter\mydefallc\fi}
\mydefallc ABCDEFGHIJKLMNOPQRSTUVWXYZeucsxht\mydefallc 

\usepackage[backend=biber,style=alphabetic,maxnames=20]{biblatex} 

\begin{document}
\begin{titlepage}
   \begin{center}
       \vspace*{2cm}

       \large{Safe Testing for Large-Scale Experimentation Platforms}
       \normalsize

       \vspace{2cm}

       \textbf{Daniel Beasley}
       
       \vspace{0.5cm}
       
       VU Thesis advisor: dr. Rianne de Heide
       
       \vspace{0.5cm}
       
       Vinted advisor: Jev Gamper
        
       \vfill
            
       MASTER THESIS
       \\
       29/08/23
            
       \vspace{0.8cm}
     
       \includegraphics[width=0.35\textwidth]{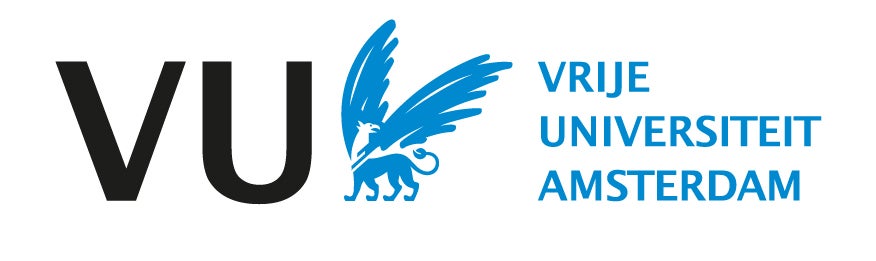}
       \includegraphics[width=0.25\textwidth]{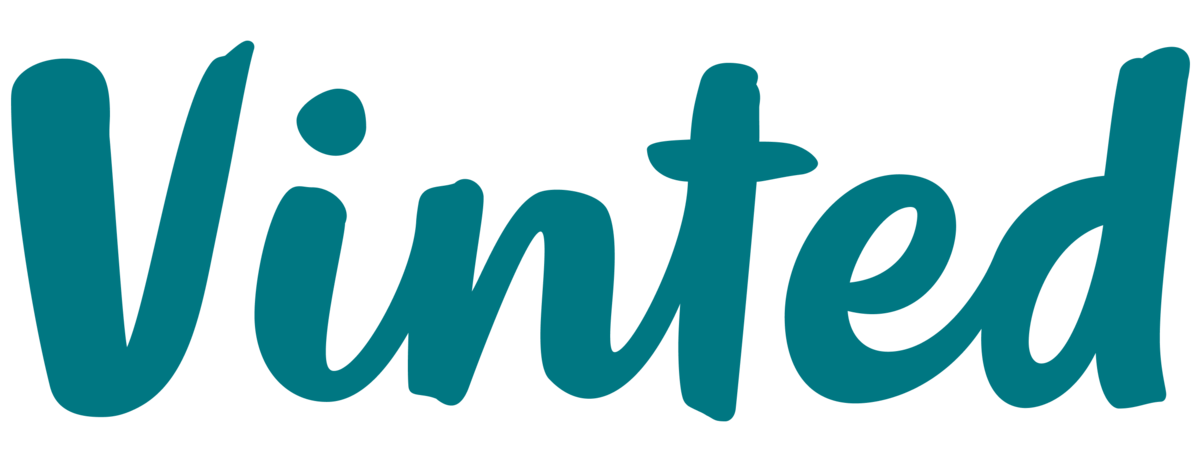}
            
       Department of Mathematics\\
       Vrije Universiteit Amsterdam\\
       Netherlands
            
   \end{center}
\end{titlepage}

\tableofcontents
\clearpage
\section{Introduction}
Randomized controlled trials (RCTs) are the gold standard for inferring causal relationships between treatments and effects. They are widely applied by scientists to deepen understanding of their disciplines. Within the past two decades, they have found applications in digital products as well, under the name A/B test. An A/B test is a simple RCT to compare the effect of a treatment (group B) to a control (group A). The two groups are compared with a statistical test which is used to make a decision about the effect.
\\
\\
Almost all statistical tests for A/B tests rely on fixed-horizon testing. This testing setup involves determining the number of users required for the test, collecting the data, and finally analyzing the results. However, this method of testing does not align with the real-time capabilities of modern data infrastructure and the experimenters' desires to make decisions quickly. Newly developed statistical methodologies allow experimenters to abandon fixed-horizon testing and analyze test results at any time. This anytime-valid inference (AVI) can lead to more effective use of experimentation resources and more accurate test results.
\\
\\
Safe testing is a novel statistical theory that accomplishes these objectives. As we will see, safe A/B testing allows experimenters to continually monitor results of their experiments without increasing the risk of drawing incorrect conclusions. Furthermore, we will see that it requires less data than standard statistical tests to achieve these results. Large technology companies are currently exploring AVI in limited capacities, but safe testing outperforms available tests in terms of the number of samples required to detect significant effects. This could lead to wide-scale adoption of safe testing for anytime-valid inference of test results.
\\
\\
This thesis contains 6 sections. Section~\ref{chapter:hypothesis} contains an introduction to hypothesis testing, as well as other statistical concepts that are relevant to the reader. It also explains how the inflexibility of classical statistical testing causes issues for practitioners. Section~\ref{chapter:safetests} introduces the concepts of safe testing. Furthermore, it derives the test statistics for the safe t-test and the safe proportion test. Section~\ref{chapter:safetestingsimulations} simulates the performance of the safe statistics and compares them to their classical alternatives. Section~\ref{chapter:msprt} compares the safe t-test to another popular anytime-valid test, the mixture sequential probability ratio test (mSPRT). Section~\ref{chapter:oce} compares the safe t-test and the mSPRT on a wide range of online experimental data. Finally, Section~\ref{chapter:vinted} is devoted to comparing the safe tests to the classical statistical tests at Vinted, a large-scale technology company.

\clearpage
\section{Hypothesis testing}\label{chapter:hypothesis}

\subsection{Introduction}

Hypothesis testing is science's method of assigning truth. Beginning with a \textit{null hypothesis} stating the current belief, the objective of a test is to determine whether or not to reject the null hypothesis \cite{lehmann2005testing}. The decision is based on the realization of a random variable $X$ with distribution $P_\theta$, where $P_\theta$ belongs to a class $\{ P_\theta : \theta \in \Theta\}$. This hypothesis class is divided into regions which accept or reject the hypothesis. For a null hypothesis $\mathcal{H}_0$, we let $\Theta_0$ represent the subset of $\Theta$ for which $\mathcal{H}_0$ is true, and $\Theta_1$ be the subset for which $\mathcal{H}_0$ is false. This is equivalent to writing $\theta \in \Theta_0$ and $\theta \in \Theta_1$, respectively. There may be an \textit{alternative hypothesis} $\mathcal{H}_1 = \{P_\theta : \theta \in \Theta_1\}$ to the null hypothesis $\mathcal{H}_0$. In the Bayesian formulation of hypothesis testing there is always an alternative $\mathcal{H}_1$, though it is not required in all frequentist formulations. If $\Theta_1 = \{\theta_1\}$, a single point, then $\mathcal{H}_1$ is called \textit{simple}. Conversely, if $|\Theta_1| > 1$, then $\mathcal{H}_1$ is \textit{composite}. The notation $\mathcal{H}_0: \theta = \theta_0$ is a condensed way of writing $\mathcal{H}_0 = \{P_\theta : \theta = \theta_0\}$. 
\\
\\
The result of a test comes from the decision function $\delta(X)$ which can take value $d_0$ to accept $\mathcal{H}_0$ or value $d_1$ to reject $\mathcal{H}_0$. Since the problem of hypothesis testing is stochastic, there always exists the possibility of committing one of two errors. The first type of error occurs when the null hypothesis is rejected when it is true, i.e. $\delta(X) = d_1$ for some $\theta \in \Theta_0$.  This is known as a \textit{Type I error} or false positive. A statistical test bounds the Type I error probability by the \textit{significance level} $\alpha$ which represents the maximum probability that this error has occurred. Mathematically, this is written as
$$P_\theta\{\delta(X) = d_1\} \le \alpha \quad \forall \theta \in \Theta_0.$$
The second type of error occurs when the hypothesis is not rejected when it is false: $\delta(X) = d_0$ when $\theta \in \Theta_1$. This error is known as a \textit{Type II error}, or false negative result. For any classical statistical test, this will also have a nonzero probability $\beta$:
$$P_\theta\{\delta(X) = d_0\} \, =\, \beta \quad \forall \theta \in \Theta_1.$$
The quantity $1-\beta$ is known as the \textit{power} of the test. This is the probability that the test correctly rejects the hypothesis in the case that it is false. For a given $\alpha$, we aim to maximize the power, which depends on the sample size of the experiment.
\\
\\
The \textit{sample size} of an experiment is the number of samples that must be collected in order to make a decision. In the process of designing a classical experiment, the experimenter will usually determine the sample size in advance. This requires assessing three quantities: the significance level $\alpha$, the power $1-\beta$, and the unknown effect size $\delta$. The \textit{effect size} is the difference between the two groups of subjects, often a combination of their mean difference and their variances. The effect size may be estimated when it is an unknown quantity, or fixed to a minimum relevant effect size, for example in clinical studies. A decrease in any of the three quantities $\alpha$, $\beta$, or $\delta$ will lead to larger sample sizes for experiments, and similarly increases lead to smaller samples sizes.
\\
\\
Historically, much of statistical testing has centered on frequentist statistics, however Bayesian statistics offers invaluable techniques for learning from data. Next, we take a deeper look at the concepts of Bayesian statistics.

\subsection{Bayesian statistics}

Bayesian statistics is an approach to statistical inference that seeks to identify possible values of an underlying parameter $\theta$ based on the observed data $X$. This is in contrast to frequentist statistics, which considers the likelihood $p_\theta(X|\theta)$ of the observations based on the parameter. Bayesian statistics is founded on \textit{Bayes' formula}, and we need to define several quantities to understand this result. A \textit{prior} represents the likely values for $\theta$ before observing any data. A prior can be based on statistical properties, such as the conjugate priors, or domain knowledge of the problem in question. For $j \in \{0, 1\}$, the prior $W_j$ is a probability distribution on $\Theta_j$ for the hypothesis class $\mathcal{H}_j$. We let $p_{W_j}(\theta|X)$ denote the \textit{posterior} probability, which is the distribution of the parameter values after observing the data. The \textit{likelihood} $p_\theta(X|\theta)$ is familiar to frequentist statisticians, as it is usually the quantity they seek to maximize. Bayes' formula, then, is $$p_{W_j}(\theta|X) = \frac{p_{\theta}(X|\theta)W_j(\theta)}{p_{W_j}(X)}.$$ where $p_{W_j}(X)$ is a normalizing constant known as the \textit{marginal likelihood}, defined as $$p_{W_j}(x) = \int_{\Theta_j} p_{W_j}(x|\theta)dW_j(\theta).$$
The ratio of marginal distributions with respect to the alternative hypothesis $\mathcal{H}_1$ and the null hypothesis $\mathcal{H}_0$ is known as the \textit{Bayes factor}: 
\begin{equation}\label{eqn:bayesfactor}
\textsc{BF}_{10}(X) := \frac{p_{W_1}(X)}{p_{W_0}(X)} = \frac{p(X|\Theta_1)}{p(X|\Theta_0)}.
\end{equation}
The Bayes factor can be thought of as the amount of evidence in favour of the alternative against the null. As we will later see, Bayes factors are intricately linked with safe testing. Another important concept in this theory is that of test martingales.
\\
\subsection{Test martingales}

Martingales have been a critical concept in probability theory since their formalization by Jean Ville \cite{ville1939}. In its simplest form, a \textit{martingale} $\{X_i\}_{i \in \mathbb{N}} = (X_1, X_2, \ldots)$ is a stochastic process such that 
\begin{equation}\label{eqn:martingale}
\Ex{X_{n+1}|X_1,\ldots,X_{n}} = X_n.
\end{equation}
A \textit{supermartingale} is defined similarly, with the $=$ in Eq~\ref{eqn:martingale} replaced by a $\le$. Ville was interested in supermartingales for the purpose of statistical testing \cite{Shafer_2011}. A \textit{test supermartingale} is a nonnegative supermartingale with $\Ex{X_1} = 1$. Ville showed that for a test supermartingale $\{X_i\}_{i \in \mathbb{N}}$, the following inequality holds:
\begin{equation}\label{eqn:villeineq}
P(\sup_i X_i \geq c) \leq 1 / c .
\end{equation}
To test $\mathcal{H}_0$ with a test supermartingale at significance level $\alpha$, we apply Ville's inequality Eq~\ref{eqn:villeineq} to reject $\mathcal{H}_0$ at time $t$ if and only if $X_t \ge 1/\alpha$. This leads to an important property of test supermartingales, known as \textit{optional stopping}. For a martingale $\{X_i\}_{i \in \mathbb{N}}$ that is measurable with respect to $(\mathcal{F}_t)_{t\in \mathbb{N}}$, a \text{stopping time} is a random variable $\tau \in \mathbb{N} \cup \{\infty\}$ such that $\{\tau = t\} \in \mathcal{F}_t$. For a test to be safe under optional stopping it means that $\{X_i\}_{i \in \mathbb{N}}$ is a test supermartingale, and $\mathcal{H}_0$ is rejected if and only if $X_\tau \ge 1/\alpha$. As we will see, this property does not apply to classical statistical tests.
\\
\\
We will now continue of our discussion of hypothesis testing with the infamous $\textsc{p}$-value.

\subsection{\textsc{p}-values}
The results of statistical tests are often reported with a \textsc{p}-value. The \textsc{p}-value corresponding to the test is a random variable \textsc{p} such that for all $0 \le \alpha \le 1$ and all $P_0 \in \mathcal{H}_0$,
$$P_0(\textsc{p} \le \alpha) \, = \, \alpha.$$
The \textsc{p}-value represents how strongly the data contradict the hypothesis. A small \textsc{p}-value suggests that the data do not accurately represent the hypothesis. \textsc{p}-values are used widely throughout the sciences, but are so misunderstood that the American Statistical Association published an article on common misconceptions in order to alleviate the issues \cite{asastatement}. One popular misconception among researchers is that a \textsc{p}-value is the probability that the null hypothesis is true. Furthermore, a \textsc{p}-value less than $0.05$ is also often used as the sole justification for scientific inference. Perhaps the most egregious issue with \textsc{p}-values is "\textsc{p}-hacking," in which unfavourable data are omitted from the analysis in order to influence the results \cite{Head2015-fx}. Thus, due to their outsized importance in scientific publications, their potential for misinterpretation, and their potential for abuse, many scientists have taken a vocal stance against \textsc{p}-values \cite{Amrhein2019-ug}, with over 800 signatories calling to abolish \textsc{p}-values. 
\\
\\
This has led to increased motivation for statisticians to develop new and improved methods to analyze scientific data. One pervasive issue has been the unreliability of statistical results during the course of an experiment.

\subsection{Optional Stopping and Peeking}

\begin{figure}[!ht]
\centering
\includegraphics[scale=0.8]{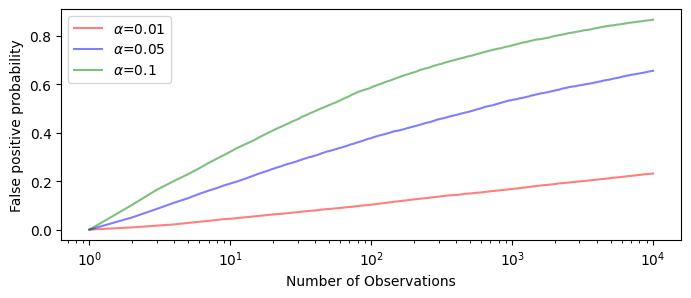}
\caption{False positive probability for the classical t-test for $\alpha = 0.01, 0.5, 0.1$ .}
\label{fig:peeking}
\end{figure}

As an experimenter conducts an A/B test, modern data infrastructure allows them to view the results in real-time. There are good reasons for them to do this. First of all, experiments are expensive to run. If an experiment’s target metric is showing negative results, there may be pressure to stop the experiment as it costs the company money. A second reason to consider stopping an experiment has to do with secondary and guardrail metrics, which provide additional information about possible causal factors of the hypothesis or unintended impacts of the test. If these metrics are showing negative results, this may suggest that the experimental feature has an unintended negative consequences for the users. A further reason for monitoring results is to check the effect size for the feature. The effect size determines the sample size, and hence the length of time that the test must run. If the effect is large, the experimenter may suggest to stop the experiment since the necessary information has been collected. 
\\
\\
Examining the results of the test before it’s complete is known as peeking, and it has unintended consequences for the results of the test. With standard A/B testing, peeking leads to an inflated false positive rate for each metric being monitored. 
Figure~\ref{fig:peeking} shows how the false positive probability increases with successive peeks. The data are derived from the same distribution and tested with a two-sided, two-sample t-test. If the data are observed at the end of the test, there should be a false positive rate of $\alpha$. However, since each peek gives a new opportunity for a false positive, the probability of a false positive becomes more and more likely throughout the test. 
\\
\\
The false positive probability increases throughout the test because the test is not safe under optional stopping. In other words, continuously monitoring the results in Figure~\ref{fig:peeking} to decide when to stop the experiment can impact the outcome of the test. This is a problem for which the ideal solution is one which allows the experimenter to monitor their results while keeping false positives below $\alpha$. As we will see, safe testing is the solution that allows for this continuous monitoring and anytime-valid inference of test results.

\subsection{Combining \textsc{p}-values and Optional Continuation}

Combining \textsc{p}-values has been a subject of debate since their origins with Pearson and Fisher \cite{Heard_2018}. These methods are often applied for meta-analysis for multiple experiments. Various methods exist for different contexts, and it is not always clear which method should be used in a given situation. Safe testing provides a simple, intuitive way to combine the results of many experiments.
\\
\\
In the section on peeking, it was mentioned that experimenters may want to make a decision about the experiment results based on an intermediate observed effect size. With traditional statistical testing, the observed results are not statistically valid, and hence correct conclusions cannot be drawn. Safe testing, however, allows the experimenter to take the decision to continue a test if more results are needed to observe a significant effect.

\subsection{A/B testing}

A/B testing at first appears as a simple application of statistical tests; however, there are nuances that are incredibly relevant to experimenters. A typical A/B test will have automated measurements of tens or possibly hundreds of metrics. Consider a test in which an experimenter wishes to measure a new feature's impact on the impact on sales on their website. The \textit{target metric} for this experiment may be total sales per user. In addition to testing the feature's impact on the total sales, they may wish to see more engagement from users that did not buy anything. This is because higher engagement with the platform can increase its value to users. Therefore, monitoring \textit{secondary metrics}, such as the number of favourited items per user, the time spent on the platform, and the proportion of searches that lead to sales may give additional information about the performance of the feature. There may, however, be unintended consequences of the feature. There may be a bug that causes the website to crash on certain browsers, or the feature may cannibalize sales of cheaper products by showing more expensive ones. It is therefore crucial to monitor so-called \textit{guardrail metrics} to ensure that the feature is working as intended.
\\
\\
Aside from the metrics in the experiment, there are other factors to consider when evaluating results. Most statistical tests assume data are independent and identically distributed. However, a new feature may attract interest from curious users, leading to unreliable metrics. This is known as the \textit{novelty effect}, and may bias the results of a test. Another point of consideration is in the time it takes for metrics to converge. Some metrics, such as the number of items viewed after a search, give instantaneous results. A metric such as the proportion of users who make a purchase may take several days to converge. This is because they may be \textit{exposed} to a test while browsing the products, and return several days later to make the purchase. This time between exposure to a test and its realization can make some metrics unreliable in the short-term.
\\
\\
A final challenge to large-scale A/B testing concerns the random assignment of users to variants. Each experiment has an associated probability for users to be assigned to either the control or test group. The results of the user's session are recorded in a database before being aggregated over the course of metric calculations. Issues in this process can lead to unequal samples in the control and test group. This is known as a sample ratio mismatch (SRM) and can indicate that the test results are biased, and therefore unreliable. It is therefore important for experimenters to continuously monitor the sample ratio of their A/B tests in order to stop erroneous experiments.
\\
\\
Having discussed A/B testing and the inflexibility of traditional statistical testing, we now introduce safe testing and how it can be applied to solve these issues.

\clearpage
\section{Safe Tests}\label{chapter:safetests}

\subsection{Introduction}
\begin{figure}[!ht]
\centering
\includegraphics[scale=0.8]{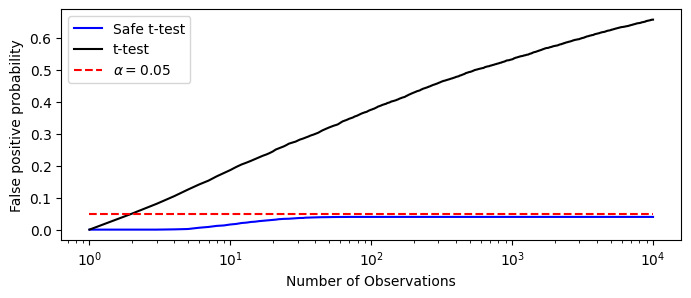}
\caption{False positive probability for the classical t-test and the safe t-test.}
\label{fig:safefalsepositive}
\end{figure}

Safe testing \cite{grünwald2023safe} is a novel method of hypothesis testing developed to address many issues with modern statistical inference. The \textit{safe} in safe testing refers to the fact that the false positive rate does not increase above $\alpha$ in the optional continuation setting. As we will see, many safe tests also allow for optional stopping \cite{grünwald2023safe}, specifically the ones we will apply to the safe t-test and the safe proportion test. Figure~\ref{fig:safefalsepositive} shows how the false positive rate of the safe t-test changes over an experiment.
\\
\\
Safe testing is based on \textit{E-variables} or (E-test statistics), which are non-negative random variables which satisfy
$$\forall P \in \mathcal{H}_0: \mathbf{E}_P\left[E\right] \le 1.$$
Under the null hypothesis, many E-variables behave as test martingales \cite{grünwald2023safe}, which are closely related to Bayes factors.
\\
\\
The original Safe Testing paper provides interpretations for E-variables based on gambling, conservative \textsc{p}-values, and Bayes factors \cite{grünwald2023safe}. The gambling analogy relates the evidence against $\mathcal{H}_0$ to monetary gain, and as such establishes an intuitive basis for interpreting the results. Suppose you are gambling at a casino with 1\$. The null hypothesis is that you will not make any money at the casino. After one round, you are left with $E\$$. According to $\mathcal{H}_0$, after $r\in \mathbb{R}^+$ rounds you should have $r\Ex{E} \le r\$$. However, if you are earning money such that $\Ex{E} > 1$, then there may be evidence to indicate that the null hypothesis is incorrect. If $E$ is sufficiently large, you may reject $\mathcal{H}_0$ altogether. 
\\
\\
In addition to this intuitive interpretation, E-variables provide many mathematical benefits as well. Earlier we highlighted \textsc{p}-values, optional stopping, and optional continuation as a few problems with classical statistical testing. We proceed now by discussing these issues in the context of E-variables.
\\
\\
The definition of E-variables leads naturally to the property of optional continuation. Suppose an experiment is conducted resulting in the E-variable $E_1$. If the experiment is continued, generating an E-variable $E_2$, then we can define a new E-variable $E^* = E_1E_2$ with expectation $\mathbf{E}_P\left[E^*\right] \le \mathbf{E}_P\left[E_1E_2\right] \le \mathbf{E}_P\left[E_1\right]\mathbf{E}_P\left[E_2\right] \le 1$. Therefore, in the context of optional continuation, one can calculate the E-variable for each segment of data and multiply them to determine the total evidence of the hypotheses. This property of E-variables is also why they are useful for meta-analyses. Combining \textsc{p}-values from multiple experiments can require a statistically-minded experimenter, while combining E-values can be done simply by multiplication.
\\
\\
The concept of statistical power isn't well defined in the optional continuation setting. This is a consequence of the power's dependency on a fixed sample size, which is in this case is random. For this reason, the concept of \textit{growth-rate optimality} (GRO) has been developed as an analogue of power in the optional continuation setting \cite{grünwald2023safe}. Optimizing an E-variable for GRO will result in its value increasing significantly with evidence against the null. Let $\mathcal{E}\left(\Theta_0\right)$ be the set of all E-variables that are test martingales under the null hypothesis. In the case of a simple alternative $\mathcal{H}_1 = \{Q\}$ with parameter $\theta_1$, GRO is defined as $$\operatorname{GRO}\left(\theta_1\right):=\sup _{E \in \mathcal{E}\left(\Theta_0\right)} \mathbf{E}_{Q}[\log E].$$ In the case that $\mathcal{H}_1$ is composite, i.e. $\mathcal{H}_1 = \{Q_{\theta_1} : \theta_1 \in \Theta_1\}$, then we seek to maximize $E$ with respect to all $Q_{\theta_1} \in \mathcal{H}_1$. These E-variables are called \textit{growth-rate optimal in the worst-case} (GROW), defined as $$\operatorname{GROW}\left(\Theta_1\right):=\sup _{E \in \mathcal{E}\left(\Theta_0\right)} \inf _{\theta \in \Theta_1} \mathbf{E}_{Q_{\theta_1}}[\log E].$$ For situations in which effect sizes are unknown or for tests with nuisance parameters, GROW may be indeterminable. However, the optimal growth can be determined relative to the unknown parameter. An E-variable with this property is known as \textit{relative GROW}. These concepts will be applied in the derivations of the safe t-statistic and the safe proportion test statistic. 
\\
\\
While there exist E-variables that are not safe under optional stopping \cite{grünwald2023safe}, A/B testing uses fairly common statistical tests for which optional stopping E-variables are available. The first such test we'll explore is the t-test, beginning with the theory behind the classical t-test.

\subsection{Classical t-test}
The t-test is a statistical test for determining whether the means of two groups of data differ significantly. Variations of the t-test differ according to the number of samples for the test and the differences in variances, so we will limit the discussion to two-sample tests as they are the object of A/B tests. Let $X_1,\ldots,X_n$ be a sample of $n$ points from a distribution $P$ with mean $\bar{X}_n = \frac{1}{n}\sum_{i=1}^n X_i$ and unknown variance. Given a second sample $Y_1,\ldots,Y_m$ of $m$ points from a distribution $Q$, a two-sample t-test determines whether the means $\bar{X}_n$ and $\bar{Y}_m$ differ significantly. The sample variances are represented as $s_{X_n}^2$ and $s_{Y_m}^2$, calculated as
$$s_{X_n}^2 = \frac{1}{n-1}\sum_{i=1}^n (X_i - \bar{X_n})^2,$$
with $s_{Y_m}^2$ determined similarly. If the two variances are assumed to be different, then one conducts a Welch's t-test, with test statistic $t$:
$$t = \frac{\bar{X}_n - \bar{Y}_m}{\sqrt{\frac{s_{X_n}^2}{n}+\frac{s_{Y_m}^2}{m}}}.$$
The t-statistic in converted to a \textsc{p}-value using the t-distribution with $\nu = n+m-2$ degrees of freedom, $$\textsc{p}(t)=\frac{2\Gamma\left(\frac{\nu+1}{2}\right)}{\sqrt{\nu \pi} \Gamma\left(\frac{\nu}{2}\right)}\bigintsss_{-\infty}^{|t|}\left(1+\frac{t^2}{\nu}\right)^{-(\nu+1) / 2},$$ which is then used to make a decision about the hypothesis. 
\\
\\
The sample size for the t-test is determined by $\alpha$, $\beta$, and the effect size $\delta$. Before the data are collected, the effect size is unknown and must be estimated. After the test, the effect size can be calculated with Cohen's d, which represents the overall difference between the groups: $$d = \frac{\bar{X_n} - \bar{Y_m}}{\sqrt{\frac{(n-1)s_{X_n}^2 + (m-1)s_{Y_m}^2}{n+m-2}}} = \frac{\bar{X_n} - \bar{Y_m}}{s_p},$$ where $s_p$ is the pooled standard deviation. 
\\
\\
With the formulation for the classical t-test established, let's examine the safe t-test.

\subsection{Safe t-test}

The safe statistic for the two-sided t-test is derived in the online appendix of Informed Bayesian T-Tests \cite{gronau2018informedappendix} and explained here for clarity. Let $D = (X_1, \ldots X_n, Y_1, \ldots, Y_m)$ be the observed data. Then the Bayes factor Eq~\ref{eqn:bayesfactor} is $$\textsc{BF}_{10}(D) = \frac{p(D|\mathcal{H}_1)}{p(D|\mathcal{H}_0)}.$$
The t-test has two free parameters, the mean difference $\mu$ and the variance $\sigma^2$. We let $\delta = \mu/\sigma$ be the effect size, where $\delta \in \Delta$. Then our hypotheses are $\mathcal{H}_0=\left\{P_{\theta_0} : \theta_0 \in \Theta_0\right\}$ where $\theta_0 = \sigma$ and $\mathcal{H}_1=\left\{Q_{\theta_1} : \theta_1 \in \Theta_1 \right\}$ where $\theta_1 = (\delta, \sigma)$. Setting priors $W_0(\theta_0)$ and $W_1(\theta_1)$, we can rewrite the Bayes factor as $$\textsc{BF}_{10}(D) = \frac{p(D|\mathcal{H}_1)}{p(D|\mathcal{H}_0)} = \frac{\int_{\Delta}\int_{\Theta_1} p_{W_1}(D|\theta, \delta)dW_1(\theta)}{\int_{\Theta_0} p_{W_0}(D|\theta)dW_0(\theta)}.$$
Following the procedure in \cite{pérezortiz2022estatistics}, we can set the prior $W_0$ to be the right Haar prior, $W_0(\sigma) \propto \sigma^{-1}$. Furthermore, we let the alternative prior $W_1(\delta, \sigma) = W_0(\sigma)\pi(\delta)$ for the two point prior $\pi(\delta)$. The marginal distribution for $\mathcal{H}_0$ becomes:
$$p\left(D|\mathcal{H}_0\right)=\frac{\Gamma\left(\frac{\nu+1}{2}\right)}{2 \pi^{\frac{\nu+1}{2}}(\nu+2)^{\frac{1}{2}}}\left(n_\delta\left[\bar{X}_n-\bar{Y}_m\right]^2+\nu s_p^2\right)^{-\frac{\nu+1}{2}},$$
where $\nu = n + m - 2$ is the degrees of freedom, $n_\delta = (1/n + 1/m)^{-1}$ is the effective sample size and $\nu s_p^2 = (n-1)s_{X_n}^2 + (m-1)s_{Y_m}^2$ is the pooled sum of squares. The marginal distribution for $\mathcal{H}_1$ is determined by integrating with respect to $\delta$ and $\sigma$, giving \cite{gronau2018informedappendix}
\begin{equation}\label{eqn:twosidedmess}
p(D | \mathcal{H}_1)=2^{-1} \pi^{\frac{1-n}{2}} n^{-\frac{1}{2}}\left(n_\delta\left[\bar{X}_n-\bar{Y}_m\right]^2+\nu s_p^2\right)^{-\frac{\nu+1}{2}} e^{-\frac{n_\delta}{2} \delta^2}[A(D | \mathcal{H}_1)+B(D | \mathcal{H}_1)],
\end{equation}
where
$$
\begin{aligned}
& A(D | \mathcal{H}_1)=\Gamma\left(\frac{\nu+1}{2}\right) {}_1 F_1\left(\frac{\nu+1}{2} ; \frac{1}{2} ; \frac{n_\delta^2\left[\bar{X}_n-\bar{Y}_m\right]^2 \delta^2}{2\left(\nu s_p^2+n_\delta\left[\bar{X}_n-\bar{Y}_m\right]^2\right)}\right) \\
& B(D | \mathcal{H}_1)=\frac{\sqrt{2} \delta \left(n_\delta\left[\bar{X}_n-\bar{Y}_m\right]\right)}{\sqrt{\nu s_p^2+n_\delta\left[\bar{X}_n-\bar{Y}_m\right]^2}}\Gamma\left(\frac{\nu+2}{2}\right) {}_1 F_1\left(\frac{\nu+2}{2} ; \frac{3}{2} ; \frac{n_\delta^2\left[\bar{X}_n-\bar{Y}_m\right]^2 \delta^2}{2\left(\nu s_p^2+n_\delta\left[\bar{X}_n-\bar{Y}_m\right]^2\right)}\right),
\end{aligned}
$$
and ${}_1F_1(a; b; z)$ is the confluent hypergeometric function. To simplify Eq~\ref{eqn:twosidedmess}, first we recognize that $B(D| \mathcal{H}_1)$ is odd, and so it must equal zero for the two-sided t-test. Let $t$ denote the classical t-statistic $t = \frac{[\bar{X_n} - \bar{Y_m}]}{s_p/\sqrt{n_\delta}}$ and let $a = \frac{t^2}{\nu + t^2}$. Then the Bayes factor is equal to $$\textsc{BF}_{10}(D) = \frac{p(D|\mathcal{H}_1)}{p(D|\mathcal{H}_0)} = e^{-\frac{n_\delta}{2} \delta^2}{}_1 F_1\left(\frac{\nu+1}{2} ; \frac{1}{2} ; \frac{a n_\delta \delta^2}{2}\right).$$
Finally, let $z = -a n_\delta \delta^2 / 2$ and apply Kummer's transformation ${}_1F_1(a; b; z) = e^z{}_1F_1(b-a; b; -z)$ to get the E-variable
$$
\begin{aligned}
E &= e^{(1-a^{-1})z}{}_1F_1\left(\frac{-\nu}{2};\frac{1}{2};z\right).
\end{aligned}
$$
The one-sided safe t-test statistic has been shown to be GROW and the two-sided test statistic to be relative GROW \cite{pérezortiz2022estatistics}. Next, we discuss the $\chi^2$ test and its safe alternative.

\subsection{\texorpdfstring{$\chi^2$} --test}

The $\chi^2$ test is a classical statistical test that is used to assess the distribution of contingency table cells. A contingency table contains the frequencies of the multinomial data, allowing one to assess the similarities of the two distributions' parameters. In the case of binomial data, the contingency table is 2x2, which will be the focus of this section.
\\
\\
Let $\{g_{1,i}\}_{i \in \mathbb{N}}$ and $\{g_{2,i}\}_{i \in \mathbb{N}}$ be two sequences of Bernoulli data: $g_{j,i} \sim \text{Bernoulli}(\theta_j)$ for $\theta_j \in \{0, 1\}$ and $j \in \{1, 2\}$. Table~\ref{contingency_table} shows a contingency table of the agreement of the sequences. For example, the cell labelled "$a$" shows the total number of times $g_{i, 1} = g_{i, 2} = 0$.
\begin{table}[h!]
\centering
\begin{tabular}{ |c|c|c| } 
 \hline
 Example 2x2 table & $g_{2,i} = 0$ & $g_{2,i} = 1$ \\ 
 \hline
 $g_{1,i} = 0$ & a & b \\ 
 \hline
 $g_{1,i} = 1$ & c & d \\ 
 \hline
\end{tabular}
\caption{Example contingency table for comparing two sequences with $g_{1, i}, g_{2, i} \in \{0, 1\}$}
\label{contingency_table}
\end{table}
\\
The contingency table can be used to test a hypothesis about the joint distribution of $\{g_{1,i}\}_{i \in \mathbb{N}}$ and $\{g_{2,i}\}_{i \in \mathbb{N}}$. The $\chi^2$ statistic is computed by comparing the values of a contingency table with their expected value under a given $\mathcal{H}_0$. Letting $E_i$ be the expected value for cell $i$ under $\mathcal{H}_0$ and $O_i$ be the observed count for cell $i$, the $\chi^2$ test statistic is $$\chi^2 = \sum_{i} \frac{(O_i-E_i)^2}{E_i}.$$
The $\chi^2$ statistic in converted to a \textsc{p}-value using the $\chi^2$ distribution with $(r-1)(c-1)$ degrees of freedom, where $r$ and $c$ are the number of rows and columns in the table. As with the classical t-test, the $\chi^2$ is not safe under optional stopping, and thus peeking can inflate their false positive rate \cite{twitter_savi}. For this reason, safe alternatives that allow anytime-valid inference have been developed, which we will explore now.

\subsection{Safe Proportion Test}

With the classical theory established, we can develop the safe proportion test statistic \cite{Turner_2019}. This derivation of the relative GROW test statistic closely follows \cite{turner2022generic}. We begin with a null hypothesis $\mathcal{H}_0=\left\{P_{\theta_a, \theta_b}:\left(\theta_a, \theta_b\right) \in \Theta_0\right\}$, where $\Theta_0=\{(\theta, \theta): \theta \in \Theta\}$, and a simple alternative hypothesis $\mathcal{H}_1=\left\{P_{\theta_a^*, \theta_b^*}:\left(\theta_a^*, \theta_b^*\right) \in \Theta_1\right\}$ for some fixed $(\theta_a^*, \theta_b^*) \in \Theta^2$. We observe $n_a$ outcomes in group $a$ and $n_b$ outcomes in group $b$, with $n := n_a + n_b$. Let $y_a^{n_a}=\left(y_{1, a}, \ldots, y_{n_a, a}\right) \in \mathcal{Y}^{n_a}$ and $y_b^{n_b}=\left(y_{1, b}, \ldots, y_{n_b, b}\right) \in \mathcal{Y}^{n_b}$. Then the marginal probability density of the observations is given by $$p_{\theta_a, \theta_b}\left(y_a^{n_a}, y_b^{n_b}\right):=p_{\theta_a}\left(y_a^{n_a}\right) p_{\theta_b}\left(y_b^{n_b}\right)=\prod_{n=1}^{n_a} p_{\theta_a}\left(y_{n, a}\right) \prod_{n=1}^{n_b} p_{\theta_b}\left(y_{n, b}\right).$$
Next, consider the quantity
\begin{equation}\label{eqn:2x2evariable}
\begin{aligned}
& s\left(y_a^{n_a}, y_b^{n_b} ; n_a, n_b, \theta_a^*, \theta_b^*\right):= \\
& \frac{p_{\theta_a^*}\left(y_a^{n_a}\right)}{\prod_{i=1}^{n_a}\left(\frac{n_a}{n} p_{\theta_a^*}\left(y_{i, a}\right)+\frac{n_b}{n} p_{\theta_b^*}\left(y_{i, a}\right)\right)} \cdot \frac{p_{\theta_b^*}\left(y_b^{n_b}\right)}{\prod_{i=1}^{n_b}\left(\frac{n_a}{n} p_{\theta_a^*}\left(y_{i, b}\right)+\frac{n_b}{n} p_{\theta_b^*}\left(y_{i, b}\right)\right)}.
\end{aligned}
\end{equation}
By Theorem 1 of \cite{turner2022generic}, the random variable $S_{\left[n_a, n_b, \theta_a^*, \theta_b^*\right]}:=s\left(Y_a^{n_a}, Y_b^{n_b} ; n_a, n_b, \theta_a^*, \theta_b^*\right)$ is an $\textsc{E}$-variable. To extend the process to composite $\mathcal{H}_1$, we set a prior $W_1$ on the alternative $\Theta_1 \subseteq \Theta^2$. Next, define $p_{W_1, a}(y):=\int p_{\theta_a}(y) d W_1\left(\theta_a\right)$ and $p_{W_1, b}(y):=\int p_{\theta_b}(y) d W_1\left(\theta_b\right)$ as the marginal posterior densities. Then the expression Eq~\ref{eqn:2x2evariable} becomes
\begin{equation}\label{eqn:2x2prior}
\begin{aligned}
& s\left(y_a^{n_a}, y_b^{n_b} ; n_a, n_b, W_1 \right) := \\
& \frac{\prod_{i=1}^{n_a} p_{W_{1, a}}\left(y_{i, a}\right)}{\prod_{i=1}^{n_a}\left(\frac{n_a}{n} p_{W_{1, a}}\left(y_{i, a}\right)+\frac{n_b}{n} p_{W_{1, b}}\left(y_{i, a}\right)\right)} \cdot \frac{\prod_{i=1}^{n_b} p_{W_{1, b}}\left(y_{i, b}\right)}{\prod_{i=1}^{n_b}\left(\frac{n_a}{n} p_{W_{1, a}}\left(y_{i, b}\right)+\frac{n_b}{n} p_{W_{1, b}}\left(y_{i, b}\right)\right)}.
\end{aligned}
\end{equation}
The random variable $S_{\left[n_a, n_b, \theta_a^*, \theta_b^*\right]}$ can be constructed sequentially, using the property that a product of two $\textsc{E}$-variables is again an $\textsc{E}$-variable. First, assume the data come in batches $Y_{(1)}, Y_{(2)}, \ldots$ with $n_a$ outcomes in group $a$ and $n_b$ outcomes in group $b$. Then, with a prior $W_1$ on batch $Y_{(1)}$, we can use the posterior $W_1 | Y^{(j-1)}$ as a prior for batch $j$. Then the $\textsc{E}$-variable after $m$ batches is
\begin{equation}\label{eqn:2x2sequential}
\begin{aligned}
S_{\left[n_a, n_b, W_1\right]}^{(m)}:=\prod_{j=1}^m S_{(j),\left[n_a, n_b, W_1\right]} ; \quad S_{(j),\left[n_a, n_b, W_1\right]}:=s\left(Y_{(j)} ; n_a, n_b, W_1 | Y^{(j-1)}\right).
\end{aligned}
\end{equation}
where $s\left(Y_{(j)} ; n_a, n_b, W_1 | Y^{(j-1)}\right)$ is calculated using $W_1 | Y^{(j-1)}$ as a prior in Eq~\ref{eqn:2x2prior}. For a sample space $\mathcal{Y} = \{0, 1\}$ we apply a binomial model such that $Y_{a} \sim \text{Bin}(n_a, \theta_a)$ and $Y_{b} \sim \text{Bin}(n_b, \theta_b)$. Let $n=n_a+n_b$ and let $n_{a1}$, $n_{b1}$ denote the number of $1$s in $Y_a$ and $Y_b$. Under $\mathcal{H}_0$, the joint distribution of $y_a^{n_a}$ and $y_b^{n_b}$ becomes $$p_{\theta_0}\left(y_a^{n_a}, y_b^{n_b}\right):=\theta_0^{n_1}\left(1-\theta_0\right)^{n-n_1} \text{, where } \theta_0 = \frac{n_a}{n}\theta_a + \frac{n_b}{n}\theta_b$$ and $n_1 = n_{a1} + n_{b1}$. Under $\mathcal{H}_1$ the joint distribution is 
\begin{equation}\label{eqn:2x2bernoulli}
\begin{aligned}
p_{\theta_a, \theta_b}\left(y_a^{n_a}, y_b^{n_b}\right) = \theta_a^{n_{a 1}}\left(1-\theta_a\right)^{n_a-n_{a 1}} \theta_b^{n_{b 1}}\left(1-\theta_b\right)^{n_b-n_{b 1}}.
\end{aligned}
\end{equation}
Combining \ref{eqn:2x2prior}, \ref{eqn:2x2sequential}, and \ref{eqn:2x2bernoulli} and simplifying (see \cite{turner2022generic} for details) gives the final expression for the relative GROW $\textsc{E}$-variable of batch size $n_a + n_b$:
\begin{equation}\label{eqn:2x2teststatistic}
S_{(j), \left[n_a, n_b, W_1\right]} = 
\frac{{\theta_a^{{n_{a1}}} (1 - \theta_a)^{n_{a} - n_{a1}}} \cdot \theta_b^{{n_{b1}}} (1 - \theta_b)^{{n_{b}} - {n_{b1}}}}{{\theta_0^{{n_{a1}} + {n_{b1}}} (1 - \theta_0)^{({n_{a}} - {n_{a1}}) + ({n_{b}} - {n_{b1}})}}}.
\end{equation}
\\
In the next section, we compare the safe t-test and the safe proportion test to their classical alternatives.

\clearpage
\section{Safe Testing Simulations}\label{chapter:safetestingsimulations}

\subsection{Introduction}
In this section, we compare the classical t-test with the safe t-test, and the $\chi^2$ test with the safe proportion test. A thorough library for safe testing has been developed in R \cite{ly2020}. With the goal of increasing adoption in the field of data science, we ported the code for the safe t-test and the safe proportion test into Python. 
\subsection{Python Implementation}

While the logic of the safe t-test remains the same, there were a number of inefficiencies in the original code that needed to be addressed in order to work with large sample sizes. The improvements are detailed here.
\\
\\
The first improvement comes in determining the sample size required for a batch process of the data. The original function performs a linear search from $1$ to an arbitrary high number. For each possible sample size in the range, the function calculates the E-value based on the sample sizes, degrees of freedom, and the effect size. The loop breaks when the E-value is greater than $1/\alpha$. Since this is a monotonically increasing function, a binary search speeds up the calculation considerably, reducing the computational complexity from $\mathcal{O}(n)$ to $\mathcal{O}(\log{n})$. This optimization proved to be necessary when working with millions of samples.
\\
\\
The next speed improvement necessary is calculating the stopping time for a power of $1-\beta$. This is determined through simulation of data differing by the minimal effect size. Over the course of $N$ simulations, data of length $m$ are individually streamed to determine the point at which the E-value crosses $1/\alpha$. Once again, this process is done through a linear search. To optimize this function, the calculation of the martingale is parallelized over the whole vector of length $m$. The computational complexity remains $\mathcal{O}(Nm)$, but the vector computation takes place in Numpy code, as opposed to a Python loop. Numpy code is written in C, hence the calculation is much faster.
\\
\\
The final modification is not in reducing computational complexity, but in improving the capabilities of the safe proportion test. This test was written in R as a two-sample test with fixed batch sizes. For our use case, a one-sample test with variable batch sizes was required to detect sample mismatch ratio, and was therefore developed for the Python package.

\subsection{Comparing the t-test with the Safe t-test}

\begin{figure}[!ht]
\centering
\includegraphics[scale=0.6]{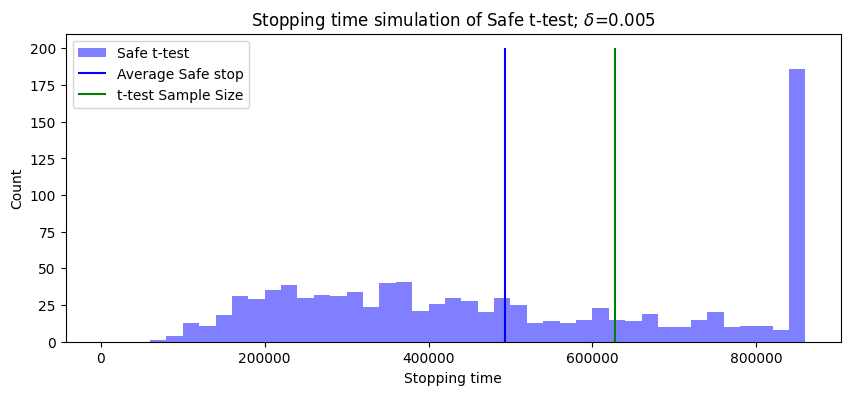}
\caption{Histogram of stopping times for the safe t-test. The solid vertical line shows the average stopping time for the safe t-test and the classical t-test.}
\label{fig:stopping}
\end{figure}

\begin{figure}[!ht]
\centering
\includegraphics[scale=0.7]{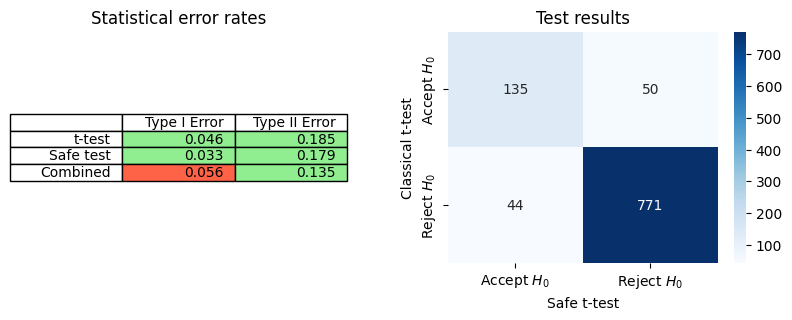}
\caption{Statistical error rates for both the safe and classical t-tests, and the results from combining their decisions (left); Decisions of the safe and classical t-tests on 1000 simulations (right).}
\label{fig:error_rates}
\end{figure}

\begin{figure}[!ht]
\centering
\includegraphics[scale=0.7]{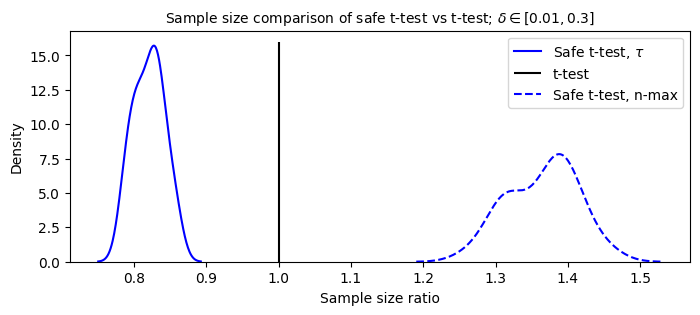}
\caption{Density histogram of the ratio of safe t-test sample sizes to the classical t-test.}
\label{fig:sample_sizes}
\end{figure}

The most straightforward way to understand the safe t-test is to compare it with its classical alternative. We perform simulations of an effect size $\delta$ and a null hypothesis $\mathcal{H}_0:\delta=0$. Setting the significance level $\alpha=0.05$ we can simulate an effect size $\delta$ between two groups to determine when the test is stopped. If the simulated E-value crosses $1/\alpha = 20$, the test is stopped with $\mathcal{H}_0$ rejected. If no effect is detected, the test is stopped at a power of $1-\beta = 0.8$, as this power is common within industry. Figure~\ref{fig:stopping} shows simulations of stopping times and decisions of the safe test compared to the t-test.
\\
\\
As we can see from the average stopping times in Figure~\ref{fig:stopping}, the safe t-test uses fewer than 500,000 samples to deliver statistically valid results, while the classical t-test requires over 600,000. However, the sample size required to reach $1-\beta$ power for the safe t-test is approximately 850,000, much larger than that of the classical t-test. One may ask whether it is acceptable to simply conduct the safe t-test until the classical t-test sample size. Figure~\ref{fig:error_rates} (left) shows the impact of this action on the statistical errors. By the completion of the test, both the classical t-test and the safe t-test are meeting the requirement that the Type I errors are below $\alpha=0.05$ and the Type II errors are below $\beta=0.2$. However, combining the two tests results in an inflated Type I error rate, and hence will not meet the experimenter's expected level of statistical significance. Given the savings in test duration, there may be motivation to develop methods combine these tests in the future such that the false positive rate remains below $\alpha$, for example using the Bonferroni correction.
\\
\\
As well as the overall conclusions of the two tests, it is interesting to consider the experiments for which the classical t-test and the safe t-test disagree. As seen in Figure~\ref{fig:error_rates} (right), while both tests achieve 80\% power, they do so in very different ways. Many simulations for which the classical t-test accepts $\mathcal{H}_0$ are rejected by the safe t-test, and vice versa. This difference in outcomes will likely be difficult to internalize for practitioners who consider the t-test to be the source of truth for their platform.
\\
\\
While Figure~\ref{fig:stopping} evaluates safe stopping times for a fixed effect size, it is important to consider the results for a wide range of effect sizes. To aggregate the results of effect sizes from $0.01$ to $0.3$, we normalize the stopping times by the t-test stopping time. The results of this analysis can be seen in Figure~\ref{fig:sample_sizes}.
\\
\\
The plot of Figure~\ref{fig:sample_sizes} shows both the average stop of the safe t-test and the sample size required for 80\% power. On average, the safe test uses 18\% less data than the t-test. In order to achieve the same power of 80\%, however, the safe test uses 36\% more data. Given that most A/B tests do not result in the rejection of $\mathcal{H}_0$ \cite{abfattails}, this could result in longer experiments overall for practitioners.

\subsection{Comparing the \texorpdfstring{$\chi^2$} --test with the safe proportion test}

\begin{figure}[!ht]
\centering
\includegraphics[scale=0.6]{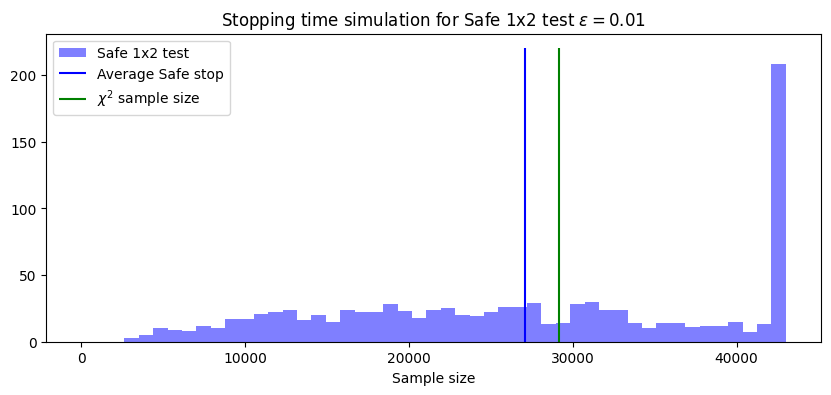}
\caption{Histogram of stopping times for the safe proportion test. The solid vertical line shows the average stopping time for the safe proportion test and the $\chi^2$ test.}
\label{fig:stopping_times_1x2}
\end{figure}

\begin{figure}[!ht]
\centering
\includegraphics[scale=0.7]{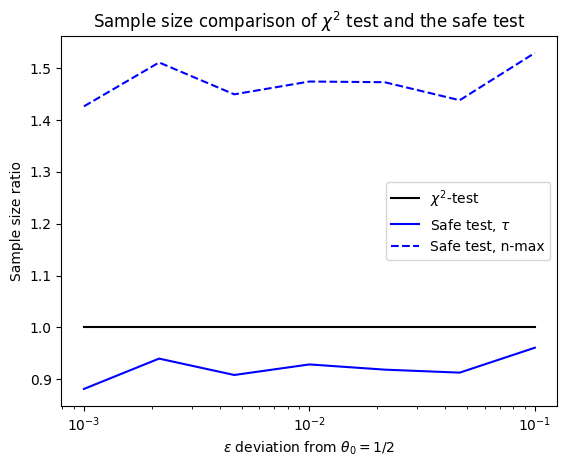}
\caption{Average and maximum stopping times of the safe proportion test  as a ratio of the $\chi^2$ sample size, with priors $\alpha_1=\beta_1 = (10\epsilon^2)^{-1}$.}
\label{fig:sample_size_1x2}
\end{figure}

As with the comparison between the t-test and the safe t-test, we can compare the $\chi^2$ test the safe proportion test. The objective of this section is to develop a test for sample ratio mismatch. A test for sample ratio mismatch will determine if the parameter of a sequence of Bernoulli trials differs from a fixed $\theta_0$ by at least $\epsilon$. For a sample $X_1, X_2, \ldots \sim \text{Bernoulli}(\theta_1)$, we wish to determine if $|\theta_1 - \theta_0| > \epsilon$. Thus, we convert the two-sided test statistic in Eq.~\ref{eqn:2x2teststatistic} to a one-sided test statistic:
$$S_{(j), \left[n_1, W_1\right]} = 
\frac{{\theta_1^{{n_{1}}} (1 - \theta_1)^{n - n_{1}}}}{{\theta_0^{{n_{1}}} (1 - \theta_0)^{{n - {n_{1}}}}}}.$$ Following the procedure in \cite{turner2022generic}, we set $W_1 \sim \text{Beta}(\alpha_1, \beta_1)$. Empirically, we've found that $\alpha_1 = \beta_1 = (10\epsilon^2)^{-1}$ works well in practice, and will therefore be used for these simulations. Furthermore, we set the power $1-\beta = 0.8$ and the significance level $\alpha=0.01$, in line with industry standards. Figure \ref{fig:stopping_times_1x2} shows the results of simulations for $\epsilon = 0.01$.
\\
\\
The results of Figure \ref{fig:stopping_times_1x2} are remarkably similar to those seen comparing the t-test and the safe t-test in Figure \ref{fig:stopping}. The safe test again uses fewer samples, on average, than its classical alternative, while the maximum stopping time to achieve the required power is higher. Next, we consider the sample sizes of the tests as a function of the difference $\epsilon$. Figure \ref{fig:sample_size_1x2} shows both the average and maximum stopping times for $\epsilon \in [0001, 0.1]$.
\\
\\
As seen in Figure \ref{fig:sample_size_1x2}, the average sample size required for the safe proportion test is less than that of the $\chi^2$ test for all values of $\epsilon$. This suggests that the safe proportion test will be competitive with the $\chi^2$ test, even for detecting small effects. Looking at these results, one may question whether it is appropriate to set a prior based on an unknown effect size. However, the prior can based the effect size calculated from the data after each sample. Hence, setting the priors based on the current effect size has no impact on the validity of the test.
\\
\\
In this section, we have compared the safe t-test and the safe proportion test with their classical alternatives. It was found that average sample sizes for the safe t-test are smaller than those of the classical t-test for a wide range of effect sizes. However, the maximum sample size can be much greater to achieve the same statistical power. Additionally, the average sample sizes of the safe proportion test are smaller than those of the $\chi^2$ test. These findings motivate further adoption of safe tests in scientific endeavours. In the next section, we compare the safe t-test to another anytime-valid test used in industry, the mixture sequential probability ratio test.

\clearpage
\section{Mixture sequential probability ratio test}\label{chapter:msprt}

\subsection{Sequential Testing}

As sophisticated A/B testing infrastructure has proliferated, so too have the opportunities to peek at test results \cite{peeking}. As we've seen, this leads to the unintended consequence of inflating the false positive rate. To take advantage of their infrastructure, then, big technology companies have begun implementing statistical methods that are valid at any time. This field of statistics is known as sequential testing, or anytime-valid inference. Sequential testing originated with Wald's seminal paper on the subject, Sequential Tests of Statistical Hypotheses \cite{wald1945}. Wald introduces the first sequential testing method, known as the sequential probability ratio test (SPRT). The SPRT is a one-sample test of size $m$ that divides the sample space into three mutually exclusive regions corresponding to the decision to be taken: either accept $\mathcal{H}_0$, reject $\mathcal{H}_0$, or continue sampling. The quantity to determine the decision is the posterior probability of the data under $\mathcal{H}_1$ divided by the posterior probability under $\mathcal{H}_0$, $P(D|\mathcal{H}_1)/P(D|\mathcal{H}_0)$. This is the well-known Bayes factor between the alternative and null hypotheses and is closely related to E-variables in safe testing \cite{grünwald2023safe}.
\\
\\
Wald and Wolfowitz proved that the SPRT is the optimal sequential test in terms of statistical power \cite{waldwolfowitz}. It should be noted, however, that their formulation of a sequential test is not aligned with that of safe tests. Their proof is based on dividing the probability ratio space into three regions: accept $\mathcal{H}_0$, reject $\mathcal{H}_0$, or continue sampling. Conversely, the safe t-test is optimal in terms of GROW \cite{pérezortiz2022estatistics}, which means that the E-variable $E$ will grow fastest when $\mathcal{H}_0$ is not true. The decision to reject $\mathcal{H}_0$ is taken when $E \ge 1/\alpha$, while the opposing decision to accept $\mathcal{H}_0$ can be taken at any time. Understand the differing formulations of these sequential tests and their optimality proofs should help to internalize the relative performances of the two tests.

\subsection{Mixture SPRT}

Developing an A/B test for sequential testing involved expanding the SPRT to function with two-sample data. This was accomplished by Johari et al. \cite{peeking} who pioneered a method of A/B testing known as the mixture Sequential Probability Ratio test (mSPRT). This test has been adopted in large technology companies such as Uber and Netflix \cite{sequential_frameworks}. As with the safe t-test, the mSPRT performs optimally with granular, sequential data. The mSPRT is essentially similar to the SPRT, with a prior belief that the true parameter lies close to $\theta_0$. Let's examine the mathematical details of this test in more depth.
\\
\\
As this is a two-sided test, we let $\mathcal{H}_0 : \theta = \theta_0$ and $\mathcal{H}_0 : \theta \ne \theta_0$ be the null and alternative hypotheses, respectively. We define $H$ to be a \textit{mixing distribution} over $\Theta$ with positive density $h$, and let $f_\theta$ be the density of the data with parameter $\theta$. The likelihood ratio of the data is defined as $$\lambda = \frac{L(\theta)}{L(\theta_0)} = \prod_{i=1}^n \frac{f_\theta(X_i)}{f_{\theta_0}(X_i)}.$$ Mixing $H$ over the parameter space $\Theta$ means applying a probability distribution to the possible values of the parameter being tested. The mSPRT test statistic is calculated by mixing the null hypothesis $\mathcal{H}_0$ with the likelihood ratio of the data:
$$\Lambda_n^{H, \theta_0} = \int_{\Theta} \prod_{i=1}^n \frac{f_\theta(X_i)}{f_{\theta_0}(X_i)}h(\theta)d\theta.$$
The central limit theorem means we can select a normal distribution for the density $f_\theta = \mathcal{N}(\theta, \sigma^2)$. For the mixing distribution $H$ it is convenient to choose a normal distribution centered at $\mathcal{H}_0$, such that $H = \mathcal{N}(\theta_0, \gamma^2)$. The hyperparameter $\gamma^2$ is known as the mixing variance, and details will follow for how to select an optimal value.
\\
\\
The assumptions of normality for the mSPRT test with data $(X_1, \ldots, X_n ; Y_1, \ldots, Y_n)$ and means $\bar{X}_n$ and $\bar{Y}_n$ leads to the test statistic
$$\tilde{\Lambda}_n^{H, \theta_0} = \sqrt{\frac{2\sigma^2}{2\sigma^2 + n\gamma^2}}\exp\Biggl\{\frac{n^2\gamma^2(\bar{Y}_n - \bar{X}_n - \theta_0)^2}{4\sigma^2(2\sigma^2 + n\gamma^2)}\Biggr\}.$$
As with the safe t-statistic, the mSPRT statistic is a test martingale \cite{peeking} and rejects $\mathcal{H}_0$ at the first time $\tau$ that $\Lambda_\tau^{H, \theta_0} \ge 1/\alpha$. In practice, however, mSPRT test statistics are usually converted to \textsc{p}-values by inverting the statistic and taking the minimal value throughout the experiment: $$p_0=1;\quad p_n = \min\{p_{n-1}, 1/\Lambda_n^{H, \theta_0}\}.$$ We will keep the mSPRT statistic in its martingale form in order to compare the performance with the safe t-test.
\\
\\
A challenge to obtaining consistent results for the mSPRT is in selecting an optimal mixing variance $\gamma^2$. This parameter is intristically linked to the effect size $\delta$ of the experiment. Liu et al. \cite{liu2022datasets} propose a method for selecting $\gamma^2$ by collating the value of Cohen's $d$ with the sample variance after $n$ samples,
$$\gamma^2 = d\cdot(s_p^2)_n,$$
where $d$ is Cohen's $d$ and $(s_p^2)_n$ is the sample variance of the first $n$ data points. This method of selecting $\gamma^2$ will be used for the following simulations.

\subsection{mSPRT and the safe t-test}

\begin{figure}[!ht]
\centering
\includegraphics[scale=0.7]{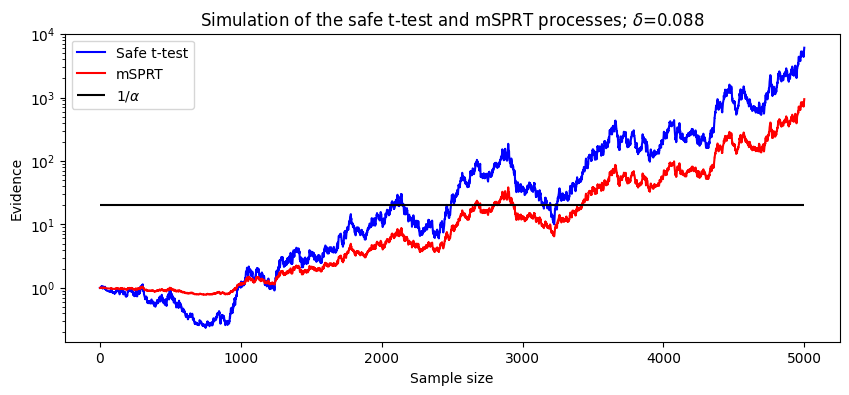}
\caption{Realizations of the safe t-test and the mSPRT statistics on normal data with effect size $\delta$.}
\label{fig:msprt_martingale}
\end{figure}

\begin{figure}[!ht]
\centering
\includegraphics[scale=0.65]{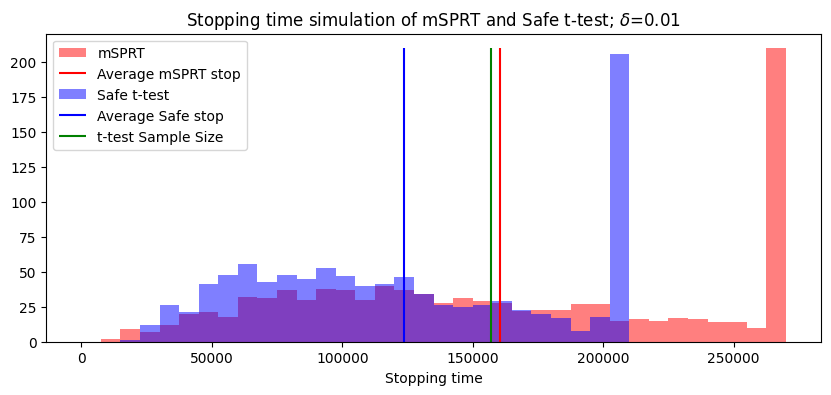}
\caption{Histogram of stopping times for mSPRT and the safe t-test. The solid vertical lines show the average stopping time for the safe t-test, the classical t-test and the mSPRT.}
\label{fig:stop_time_comparison}
\end{figure}

\begin{figure}[!ht]
\centering
\includegraphics[scale=0.65]{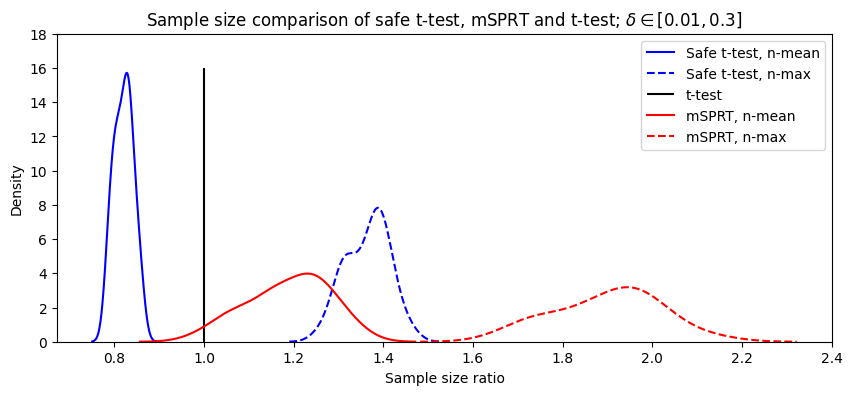}
\caption{Density histogram of average stopping times and maximum stopping times at $1-\beta$ power, for the safe t-test and the mSPRT.}
\label{fig:sample_size_comparison}
\end{figure}

\begin{figure}[!ht]
    \centering
    \includegraphics[scale=0.7]{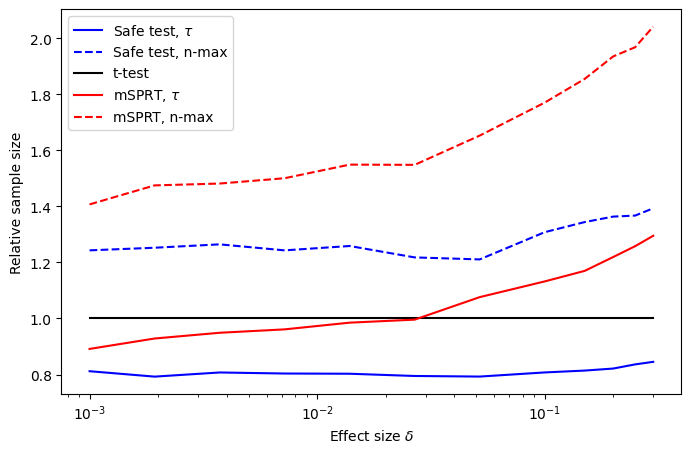}
    \caption{Average and maximum stopping times of the safe t-test and the mSPRT, as a ratio of the classical t-test sample size.}
    \label{fig:rel_sample_size}
\end{figure}

\begin{figure}[!ht]
    \centering
    \subfloat[\centering $\beta=0.2$ and $\delta=0.01$]{{\includegraphics[scale=0.4]{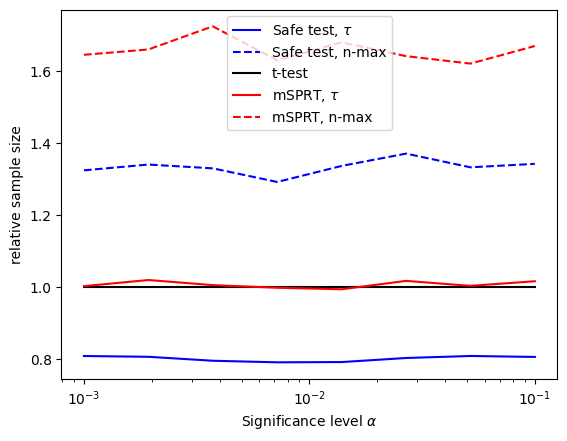} }}%
    \qquad
    \subfloat[\centering $\alpha=0.05$ and $\delta=0.01$]{{\includegraphics[scale=0.4]{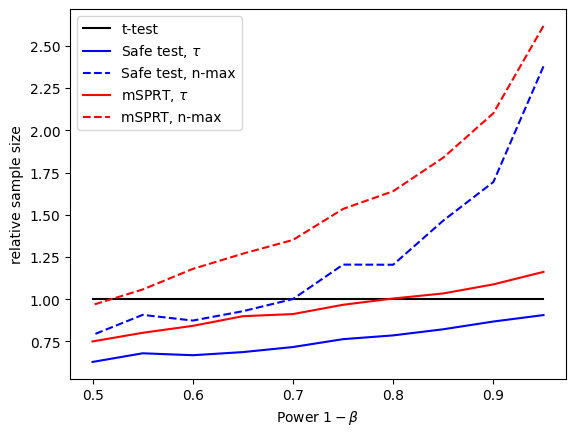} }}%
    \caption{Stopping times for the safe t-test and the mSPRT, normalized by the classical t-test, as a function of $\alpha$ (left) and $\beta$ (right).}%
    \label{fig:alphabeta}
\end{figure}

In this section, we will compare mSPRT and the safe t-test in terms of power, sample size, and other properties. 
\\
\\
We will first consider the performance of the statistics on a pair of random normal samples with a mean difference of $\delta$ and unit variance. Both of the statistics behave as test martingales, so we can compare them visually as they accumulate evidence for and against $\mathcal{H}_0$. Figure~\ref{fig:msprt_martingale} shows a simulation of these processes.
\\
\\
The first thing to notice is that both tests draw similar conclusions from the data. In the first 1000 samples, there is evidence in favour of $\mathcal{H}_0: \delta=0$ which causes both test statistics to decrease. Following evidence in favour of the alternative hypothesis $\mathcal{H}_1: \delta \ne 0$, both test statistics increase until they cross the $1/\alpha$ threshold. A second observation is in the magnitude that each test weighs evidence. As evidence initially supports the null hypothesis, the safe test statistic decreases much more quickly than the mSPRT statistic. However, as the data supporting $\mathcal{H}_1$ increases, the safe statistic surpasses the mSPRT statistic, staying far above for the remainder of the experiment. A final observation is in when the statistics cross the $1/\alpha$ threshold. While the safe test needs less than 2100 samples to reject the null, mSPRT needs over 2700 samples.
\\
\\
With an understanding of how the safe t-test and mSPRT perform on a random sample, we can now consider many simulations with the same effect size $\delta$. The goal of these tests is to stop when enough evidence against $\mathcal{H}_0: \delta=0$ has been collected. Therefore, we will compare the $1/\alpha$ stopping times of these test statistics. In cases for which no effect is detected, the test is stopped at a power of $1-\beta = 0.8$. Figure~\ref{fig:stop_time_comparison} shows the result of many simulations of this process.
\\
\\
The stopping time histogram in Figure~\ref{fig:stop_time_comparison} shows that for the simulated data with an effect size of $\delta=0.01$, safe tests are able to conclude much more quickly than mSPRT. On average, the safe test uses 22\% less data than the mSPRT. For tests that do not reach the $1/\alpha$ threshold, the test is stopped without rejecting the null hypothesis. About 20\% of the both the safe test and the mSPRT reach this threshold, as is expected for a test with 80\% power. These sample size results can be broken down further based on the result of the test. It is interesting to know, for example, the number of samples required, on average, to reject $\mathcal{H}_0$. These results can be seen in Table \ref{table:stopping_decisions}. 
\\
\begin{table}[h!]
\centering
\begin{tabular}{ |c||c|c|c| } 
 \hline
 \multicolumn{4}{|c|}{\text{Number of Samples Required}} \\
 \hline\hline
 Test Result & safe t-test & mSPRT & (ratio) \\ 
 \hline\hline
 Reject $\mathcal{H}_0$ & 103552 & 133677 & 0.775 \\ 
 Accept $\mathcal{H}_0$ & 204751 & 265464 & 0.771 \\ 
 (Either) & 123792 & 160298 & 0.772 \\ 
 \hline
\end{tabular}
\captionof{table}{Average number of samples required to either reject or accept $\mathcal{H}_0$ for both the safe t-test and the mSPRT.}
\label{table:stopping_decisions}
\end{table}
\\
The results of Table~\ref{table:stopping_decisions} are relevant for practitioners who are particularly concerned with deviations from the null hypothesis. Uber, for example, uses mSPRT to monitor outages of their platform \cite{sequential_frameworks}. Given that the safe t-test rejects $\mathcal{H}_0$ with 22\% less data, this could decrease the time to recognize outages and hence improve response time.
\\
\\
We've seen that for simulations of $\delta=0.01$ the safe test concludes using fewer samples than the mSPRT, but it remains to be seen for different effect sizes. The following experiment is conducted on 30 effect sizes ranging from 0.01 to 0.3. There are two stopping times we wish to consider: the average stopping time for each test, and the stopping time required for $80\%$ power. To contextualize these results, we can consider the sample size ratio of each of these tests with respect to the classical A/B test. Figure~\ref{fig:sample_size_comparison} shows the average and maximum stopping times of the safe t-test and the mSPRT, in terms of a sample size ratio of the classical t-test. 
\\
\\
The solid lines in Figure~\ref{fig:sample_size_comparison} represent the average stopping time of all simulations for all effect sizes. The safe test needs about 20\% fewer samples than the t-test, while the mSPRT needs about 20\% more. The dashed lines represent the maximum sample sizes required to achieve $1-\beta$ power. While the mSPRT needs about twice as many sample as the t-test to achieve this power, the safe t-test only needs about 40\% more samples.
\\
\begin{table}[h!]
\centering
\begin{tabular}{ |c||c|c|c| } 
 \hline
 \multicolumn{4}{|c|}{\text{Ratio of Samples Required; $\delta \in [0.01, 0.03]$}} \\
 \hline\hline
 Test Result & safe t-test & t-test & mSPRT  \\ 
 \hline\hline
 Reject $\mathcal{H}_0$ & 0.68 & 1.0 & 1.06 \\ 
 Accept $\mathcal{H}_0$ & 1.36 & 1.0 & 1.96 \\ 
 (Either) & 0.82 & 1.0 & 1.20 \\ 
 \hline
\end{tabular}
\caption{Average number of samples required to either reject or accept $\mathcal{H}_0$ for both the safe t-test and the mSPRT.}
\label{table:stopping_decisions_all}
\end{table}
\\
As with Table~\ref{table:stopping_decisions}, we can compare the average stopping times for the tests based on whether they reject or accept $\mathcal{H}_0$. These results, found in Table~\ref{table:stopping_decisions_all}, show that to reject $\mathcal{H}_0$, the safe t-test uses 32\% less data than the t-test, while the mSPRT uses 6\% more. This provides further evidence that the safe t-test is more efficient than the mSPRT in reaching conclusions with the same data.
\\
\\
To compare the average stopping times as a function of effect size $\delta$, we can again normalize the sample sizes by the classical t-test sample size. The results can be seen in Figure~\ref{fig:rel_sample_size}.
\\
\\
Figure~\ref{fig:rel_sample_size} shows that the safe t-statistic sample sizes are smaller than both the classical t-test and the mSPRT for all $\delta \in \left[0.001, 0.3\right]$. The capability of the safe t-test to detect these small effect sizes is a motivator for the its use in online A/B testing.
\\
\\
Until now, all simulations have been conducted with $\alpha=0.05$ and $\beta=0.2$. To assure readers that these parameters are not biasing the results, Figure~\ref{fig:alphabeta} shows the stopping times when varying these parameters.
\\
\\
It is clear that the effectiveness of the safe t-test will extend to a variety of testing scenarios based on experimenter's needs.
\\
\\
In this section, we have compared the safe t-test and the mSPRT through various simulations. It was found that the safe t-test stops earlier than the mSPRT in all simulations. This leads to smaller sample sizes and faster experimentation. We also found that the safe t-test is able to reject $\mathcal{H}_0$ with much less data than both the classical t-test and the mSPRT. In the next section, we further analyze the performance of these statistic tests on real A/B test data.

\clearpage
\section{Online Controlled Experiments}\label{chapter:oce}

As A/B testing adoption has increased, new statistical methodologies have similarly proliferated. Variance reduction techniques such as CUPED \cite{deng2013improving} and new statistical tests such as mSPRT \cite{peeking} have become standard techniques at various technology companies. The A/B testing landscape is becoming increasingly crowded with innovative methods but there isn't a benchmark with which to easily compare and contrast them. To address this issue, researchers from Imperial College London have compiled a series of datasets for online controlled experiments (OCE) \cite{liu2022datasets}. These datasets contain real data from randomized controlled trials conducted online as well as some synthesized results. Collectively known as the OCE datasets, they can be used to benchmark and compare novel methods for conducting A/B tests.
\\
\\
The OCE datasets are a taxonomy of 13 anonmyized datasets found around the internet. The datasets contain daily snapshots of four metrics measured on 78 experiments and up to three variants. The data from the experiments can be binary, integer, or real-valued, allowing for a wide range of statistical methodologies to be tested.
\\
\\
A potential use case for the OCE datasets is benchmarking optional stopping methods, as newly developed methods may have unrealistic assumptions to ensure validity of their results. The availability of daily snapshots in the datasets means that a semi-sequential approach to testing can be applied. In the paper, mSPRT is compared to the classical t-test to validate the test on the OCE datasets. Table~\ref{table:msprtoce} shows the results.
\\
\begin{table}[h!]
\tablefont
\centering
\begin{tabular}{cc|cc}
 \multicolumn{2}{c|}{\multirow{2}{*}{mSPRT vs t-test}} & \multicolumn{2}{c}{mSPRT} \\
 & & Accept $H_0$ & Reject $H_0$ \\
 \hline
\multirow{2}{4em}{t-test} & Accept $H_0$ & 264 & 30 \\
 & Reject $H_0$ & 34 & 71 \\
\end{tabular}\par
\captionof{table}{Decisions of the mSPRT and the classical t-test on the OCE datasets.}
\label{table:msprtoce}
\end{table}
\\
In this section, the safe t-test will be conducted on the collection of OCE datasets, and the results compared with both the classical t-test and the mSPRT.

\subsection{Safe t-test on OCE datasets}

In order to benchmark the performance of the safe t-test, we can compare its results with the t-test. As we've seen in Figure~\ref{fig:error_rates} (right), the two tests do not always reach the same conclusion for each set of data. However, since the t-test is the most widely used statistical test for A/B testing, it is important to contrast the results in order to understand the situations in which the results differ. Table~\ref{table:safeoce} shows the results of the t-test and the safe t-test on the collection of OCE datasets.
\\
\begin{table}[h!]
\tablefont
\centering
\begin{tabular}{cc|cc}
 \multicolumn{2}{c|}{\multirow{2}{*}{Safe t-test vs t-test}} & \multicolumn{2}{c}{Safe t-test} \\
 & & Accept $H_0$ & Reject $H_0$ \\
 \hline
\multirow{2}{4em}{t-test} & Accept $H_0$ & 228 & 48 \\
 & Reject $H_0$ & 21 & 84 \\
\end{tabular}\par
\captionof{table}{Decisions of the safe t-test and the classical t-test on the OCE datasets.}
\label{table:safeoce}
\end{table}
\\
The safe t-test detects many more effects than the classical t-test. While, in theory, the false positive rate of the safe t-test should be below $\alpha$, it seems unlikely that all of these rejections of $\mathcal{H}_0$ correspond to true effects. Following analysis of the behaviour of the E-values over the course of these experiments, we conclude that the high number of $\mathcal{H}_0$ rejections likely has to do with novelty effect. As mentioned previously, the novelty effect refers to an increased attention for the feature shortly after its release. The result is that the assumption of independent and identically distributed data is violated, with evidence against the null hypothesis to accumulate rapidly. For a fixed-sample test this is less of an issue because the distribution reverts over the course of an experiment. However, for safe tests this can cause a rejection of $\mathcal{H}_0$ before the true impact of the feature is determined. This fact is particularly relevant to practitioners seeking to implement anytime-valid statistical testing. Next, in Table~\ref{table:safemsprtoce}, we compare the safe test and the mSPRT on the OCE datasets. 
\\
\begin{table}[h!]
\tablefont
\centering
\begin{tabular}{cc|cc}
 \multicolumn{2}{c|}{\multirow{2}{*}{Safe t-test vs mSPRT}} & \multicolumn{2}{c}{Safe t-test} \\
 & & Accept $H_0$ & Reject $H_0$ \\
 \hline
\multirow{2}{4em}{mSPRT} & Accept $H_0$ & 249 & 31 \\
 & Reject $H_0$ & 0 & 101 \\
\end{tabular}\par
\captionof{table}{Decisions of the safe t-test and the mSPRT on the OCE datasets.}
\label{table:safemsprtoce}
\end{table}
\\
Unsurprisingly given the behaviour observed in Figure~\ref{fig:msprt_martingale}, the null hypotheses rejected by the mSPRT are similarly rejected by the safe t-test. However, the safe test rejects even more of the hypotheses than the mSPRT. This is likely because the safe test is more sensitive than the mSPRT and reacts more strongly to data which contradict the null hypothesis. In the next section, we continue analyzing the performance of safe tests at a large-scale tech company, Vinted.

\clearpage
\section{Vinted A/B tests}\label{chapter:vinted}

Vinted is an online marketplace for clothing and accessories. Since its inception in 2008, it has gained over 75 million users to rapidly develop into Europe's largest secondhand clothing marketplace. With such an abundance of users, it conducts a large number of A/B tests simultaneously to deliver the best experience for its users. This makes Vinted an ideal environment to evaluate the efficacy of safe tests. In this section, we apply the safe t-test and the safe proportion tests to Vinted's experiment data. The safe t-test will be compared to the classical t-test to evaluate the results of A/B tests. In addition, the safe proportion test will be compared to $\chi^2$ test as a means to detect the sample ratio mismatch of experiments.

\subsection{Safe t-test for Vinted A/B tests}

The metrics for 162 Vinted experiments from March 2023 to June 2023 will be evaluated for this analysis. We collated cumulative daily snapshots of 143 metrics, containing the metric's mean, standard deviation, and sample size for both control and test groups. Experiments with multiple variants are treated as separate tests with the same control group. The safe t-test and the classical t-test were compared across all 42115 experiment/metric combinations in this dataset. Table~\ref{table:safetvinted} shows the results of the statistical tests at level $\alpha=0.05$.
\begin{table}[ht!]
\tablefont
\centering
\begin{tabular}{cc|cc}
 \multicolumn{2}{c|}{\multirow{2}{*}{Safe t-test vs t-test}} & \multicolumn{2}{c}{Safe t-test} \\
 & & Accept $H_0$ & Reject $H_0$ \\
 \hline
\multirow{2}{4em}{t-test} & Accept $H_0$ & 92.9\% & 0.9\% \\
 & Reject $H_0$ & 3.9\% & 2.3\% \\
\end{tabular}\par
\captionof{table}{Decisions of the safe t-test and the classical t-test on Vinted A/B tests.}
\label{table:safetvinted}
\end{table}
\\
The results of Table~\ref{table:safetvinted} show that the safe t-test and the classical t-test consistently reach the same conclusion about the significance of the metrics. The 379 cases in which the safe t-test rejects an $H_0$ that the t-test does not are consistent with the simulations demonstrating that the tests do not always agree on what constitutes a significant result. The high number of 1645 cases in which the t-test rejects $H_0$ while the safe t-test does not are more concerning. The safe t-test is more sensitive when it observes data sequentially, giving more opportunities to reject $\mathcal{H}_0$. These data are aggregated on a daily level, which effectively reduces the power of the test. With more granular data, the safe t-test would detect more effects than in this group-sequential setting.
\\
\\
The mixture sequential probability ratio test (mSPRT) was conducted on the same set of experiments. The results can be found in Table~\ref{table:msprtvinted}.
\\
\begin{table}[ht!]
\tablefont
\centering
\begin{tabular}{cc|cc}
 \multicolumn{2}{c|}{\multirow{2}{*}{mSPRT vs t-test}} & \multicolumn{2}{c}{mSPRT} \\
 & & Accept $H_0$ & Reject $H_0$ \\
 \hline
\multirow{2}{4em}{t-test} & Accept $H_0$ & 93.7\% & <0.1\% \\
 & Reject $H_0$ & 5.8\% & 0.5\% \\
\end{tabular}\par
\captionof{table}{Decisions of the safe t-test and the mSPRT on Vinted A/B tests.}
\label{table:msprtvinted}
\end{table}
\\
Comparing the results of Table~\ref{table:msprtvinted} with Table~\ref{table:safetvinted} show that the mSPRT is significantly less powerful than the safe t-test. While this is in part due to the group-sequential setting, our simulation results suggest that the mSPRT is simply a less sensitive statistical test than the safe t-test.
\\
\\
Returning to the safe t-test results, it was found that the safe t-test performed significantly better on some metrics than others. Here, we will further analyze the metrics to understand why this is the case. To quantify the safe t-test's performance on a metric, we use the phi coefficient to compare its decisions with the classical t-test. The phi coefficient, also known as Matthews correlation coefficient, is used to determine the correlation of binary variables. To understand the purpose of each metric, there is a text description of its use case within Vinted's A/B testing framework. A summary of the topics in each description can be extracted with Latent Dirichlet Allocation. Latent Dirichlet allocation (LDA) is a natural language processing technique for modelling the topics from a set of documents. In this case, LDA is used to extract the topics from the metric descriptions in the form of latent vectors. We multiply the latent vectors by the phi coefficient to find the average phi coefficient for each topic. Table~\ref{table:phicoefficient} shows words that are correlated with higher and lower phi coefficients.
\\
\begin{table}[ht!]
\centering
\begin{tabular}{c|c}
 $\phi$ coefficient & Metric description topics \\
 \hline\hline
 $\phi > 0.5$ &  search, session, browsing, impressions \\
 \hline
 $0.35 \le \phi \le 0.5$ & replies, want, price, complaint \\
 \hline
 $\phi < 0.35$ & transaction, cancellation, seller, monthly \\
\end{tabular}\par
\captionof{table}{Phi coefficient correlation with metric description topics from latent Dirichlet allocation. Note that the selection of words is not random, but serves to illustrate the metrics for which the safe t-test performs optimally.}
\label{table:phicoefficient}
\end{table}
\\
In the introduction to A/B testing, it was mentioned that some metrics take much longer to be realized. This means that the data will not be independent and identically distributed across the days of the test. Examining Table~\ref{table:phicoefficient}, we see a high correlation between the performance of the safe t-test and the classical t-test on metrics involving searches, sessions, and impressions. These are all quantities that have a short time between exposure to the test and the realization of the metric. Conversely, the safe t-test does not perform well on long-term metrics involving transactions and order cancellations. Together, these results suggest that the safe t-test will perform optimally on metrics for which results are available instantaneously.

\subsection{Safe proportion test for sample ratio mismatch}

To determine the efficacy of the safe proportion test and the $\chi^2$ test in detecting sample ratio mismatch (SRM), the distributions of 195 experiments from Vinted are analyzed. The safe test is applied to daily snapshots of the distributions, while the $\chi^2$ test is applied to the distribution on the final day of the experiment. For SRM, a significance level of $\alpha=0.01$ is used to limit the number of false positives. Beta prior values of $\alpha_1, \beta_1 = 1000$ are used for the safe proportion test. The comparison of the results between the safe proportion test and the $\chi^2$ test can be seen in Table~\ref{table:safepropchisquare}.
\\
\begin{table}[!ht]
\tablefont
\centering
\begin{tabular}{cc|cc}
 \multicolumn{2}{c|}{\multirow{2}{*}{Safe proportion test vs. $\chi^2$}} & \multicolumn{2}{c}{Safe proportion test} \\
 & & Accept $H_0$ & Reject $H_0$ \\
 \hline
\multirow{2}{4em}{$\chi^2$ test} & Accept $H_0$ & 131 & 0 \\
 & Reject $H_0$ & 12 & 52 \\
\end{tabular}\par
\captionof{table}{Sample ratio mismatch detection using safe proportion test and $\chi^2$ test.}
\label{table:safepropchisquare}
\end{table}
\\
The results of Table~\ref{table:safepropchisquare} show strong agreement between the safe proportion test and the $\chi^2$ test. Further analysis was conducted on number of days required to reject $\mathcal{H}_0 : \theta = \frac{1}{2}$ for both tests. Among the 52 experiments for which both tests detected SRM, the $\chi^2$ test took an average of 6.3 days to detect SRM. The average duration for the safe proportion test to detect SRM was 8.7 days. While the safe proportion test takes longer to detect an effect, it provides experimenters with anytime-valid results, giving confidence in the test's early rejections of $\mathcal{H}_0$. The value of this trade-off is one which industry practitioners can consider when deciding between statistical tests for SRM.

\clearpage
\section{Conclusion}

The myriad issues with \textsc{p}-values and their interpretation have led statisticians to seek new methods of information discovery. Classical statistical tests are unable to suit common research practices, such as early stopping or optional continuation of experiments. This disparity is becoming more noticeable with modern technological processes that allow frequent statistical analysis of data. Statistical objects such as test martingales and Bayes factors are seeing increased adoption as safer, more intuitive methods of hypothesis testing. In this thesis, we have explored safe testing as a solution to meet the needs of practitioners. In particular, we have focused on detecting small effect sizes common to A/B testing at large-scale technology companies.
\\
\\
The safe t-test was introduced as an anytime-valid substitute for the classical t-test. It was shown that the safe t-test uses, on average, less data to reject a null hypothesis. The effectiveness of the safe t-test was demonstrated for a wide range of effect sizes, significance levels, and statistical powers. On real world data, there remain discrepancies between the effects detected by the safe t-test and the classical t-test. Novelty effects can lead to an increased number of false positives, while batch processing increases the number of false negatives. There are also considerations in the delay between test exposure time and realization. This leads us to suggest that the ideal scenario for safe t-tests in large scale experimentation platforms is in granular data that is readily available. An A/B test's target metric is often a slow metric designed to improve the overall performance of the platform of its users. Secondary and guardrail metrics can be measured and analyzed much more quickly. These metrics are ideal candidates for safe tests to continuously monitor the performance of an A/B test.
\\
\\
The performance of the safe t-test was rigorously compared to the mSPRT. Through extensive simulation and benchmarking on real datasets, it was found that the safe t-test outperforms the mSPRT in all situations. This should encourage practitioners to adopt the safe t-test as their preferred anytime-valid test.
\\
\\
The safe proportion test was also validated as an anytime-valid test for contingency tables. Through simulation it was found that the safe proportion test requires less data to reach a decision than the $\chi^2$ test, on average. The tests agree considerably on real-world data to detect sample ratio mismatch. With the benefit of taking full advantage of modern data infrastructure, adoption of the safe proportion test to detect SRM will proceed at Vinted.
\\
\\
Despite the effectiveness of safe testing, it will likely take time and persistence on behalf of its proponents before it reaches wide-scale adoption. The greatest challenge will be in educating practitioners familiar with classical statistics. The concepts introduced in this thesis require a higher level of statistical knowledge than is common to most scientists and A/B test experimenters. If a practitioner using a safe test gets a different result than the classical alternative, there is not an intuitive way to explain this difference. Conversely, it may be easier for practitioners to understand E-variables conceptually. Due to their intrinsic interpretation as evidence against the null hypothesis, E-variables seem easier to grasp than \textsc{p}-values. Given the widespread misinterpretation of \textsc{p}-values, E-variables may give experimenters a better understanding of their results.
\\
\\
While there are challenges, we remain optimistic about the future applications of safe testing. It meets the needs of experimenters who need flexible testing scenarios based on observed evidence. In addition, research into E-variables is continuing to develop, which will lead to more safe tests and better education. With packages available in both R and Python, it has become easier for practitioners to introduce safe testing in their experiments. For this reason, we believe that safe testing will proliferate as an anytime-valid testing methodology.

\clearpage
\printbibliography
\end{document}